\documentclass[aps,pra,groupedaddress,amsmath,amssymb,twocolumn,10pt,nofootinbib,floatfix,showpacs]{revtex4-1}
\usepackage{hyperref}
\usepackage{graphicx,amsthm}
\usepackage{color}
\usepackage{txfonts} %kleinere Schriftart für Formeln
\usepackage{slashbox}

\newcommand{\ket}[1]{\left\vert{#1}\right\rangle}

\newcommand{\proj}[1]{\left\vert{#1}\right\rangle \left\langle{#1}\right\vert}

\newcommand{\braccket}[3]{{\langle #1 | #2 | #3 \rangle}}

\newcommand{\etaD}{\eta_d}

\newcommand{\DP}{\textit{Deutsch et al.}~}
\newcommand{\DurP}{\textit{D\"{u}r et al.}~}

\newcommand{\tauD}{\tau^\textrm{D}}
\newcommand{\tauDur}{\tau^\textrm{D\"{u}r}}

\newcommand{\kvec}{\vec{k}}

\begin{document}

\newtheorem{thm}{Theorem}
\newtheorem{lem}{Lemma}
\newtheorem{defi}{Definition}

\title{Quantum repeaters and quantum key distribution: the impact of entanglement distillation on the secret key rate}
\author{Sylvia Bratzik}
% \email{bratzik@thphy.uni-duesseldorf.de}
\author{Silvestre Abruzzo}
\author{Hermann Kampermann}
\author{Dagmar Bru{\ss}}
\affiliation{Institute for Theoretical Physics III, Heinrich-Heine-Universit\"at D\"usseldorf, 40225 D\"usseldorf, Germany.}

\date{\today}

\begin{abstract}
We investigate quantum repeaters in the context of quantum key distribution. We optimize the secret key rate per memory per second with respect to different distillation protocols and distillation strategies. For this purpose, we also derive an analytical expression for the average number of entangled pairs created by the quantum repeater, including classical communication times for entanglement swapping and entanglement distillation. We investigate the impact of this classical communication time on the secret key rate.  We finally study the effect of the detector efficiency on the secret key rate.
\end{abstract}

%Quantum communication, 03.67.Hk
%Quantum cryptography, 03.67.Dd
%Quantum information, 03.67.-a:
	%entanglement production, 03.67.Bg
	%optical implementations, 42.50.Ex

\maketitle
%\tableofcontents

\section{Introduction and motivation}

Losses in the optical fiber limit the distance for the distribution of entangled photon pairs and hence the range of quantum key distribution. Recent experiments cannot reach more than a few hundred kilometers (see, e.g., \cite{Stucki2009}). To overcome this problem the concept of a quantum repeater was developed \cite{Briegel1998,Dur1999}, which acts like a ``distance-amplifier'': it permits to enhance the probability that an entangled pair is created at a certain distance (see, e.g., calculations in \cite{Scarani2009}). For a recent review on quantum repeaters see \cite{Sangouard2011}. The main ingredients of a quantum repeater are entanglement swapping \cite{Zukowski1993} and entanglement distillation \cite{Bennett1996,Bennett1996a,Deutsch1996}. After the distribution of entangled photon pairs between two distant parties, one can perform quantum key distribution (for reviews see, e.g., \cite{Gisin2002,Scarani2009}).

Since the original proposal of the quantum repeater, existing protocols were analyzed or improved, inter alia \cite{Childress2006,Brask2010,Aghamalyan2011,Collins2007,Dorner2008,Jiang2009,Jiang2007a,Minar2011,Munro2008,Sangouard2007,Sangouard2008a,Simon2007,VanLoock2006,VanLoock2008,Zhao2010}. Moreover, new protocols like, e.g., the hybrid quantum repeater \cite{VanLoock2006} or quantum repeaters with atomic ensembles \cite{Duan2001} were introduced.

Recently, the following analyses of the secret key rate in connection with a quantum repeater were performed: in \cite{Scherer2011} a quantum key distribution (QKD) setup with one repeater node and without distillation is investigated. In this case, the parameters for the optimal secret key rate are explored. In \cite{Amirloo2010} the secret key rate for one node for the Duan-Lukin-Cirac-Zoller (DLCZ)-repeater \cite{Duan2001} is analyzed. Reference~\cite{Piparo2012} treats a variation of the DLCZ-repeater, namely \cite{Sangouard2007}. In \cite{Abruzzo2012} secret key rates for the original quantum repeater \cite{Briegel1998}, for the hybrid quantum repeater \cite{VanLoock2006} and for a variation of the DLCZ-repeater \cite{Minar2011} are investigated,  where distillation was considered only before the first entanglement swapping. Here, we want to lift this restriction and allow distillation in all nesting levels. 

The main goal of the current work is to analyze the achievable secret key rate under different distillation protocols and strategies. For the distillation protocols we consider a recurrence protocol \cite{Deutsch1996} and the entanglement pumping protocol \cite{Dur1999}. The protocol \cite{Deutsch1996} is more efficient regarding the final fidelity for perfect gates but at an expense of an exponentially growing number of memories. The protocol in Ref.~\cite{Dur1999} reaches a higher fidelity than the protocol in Ref.~\cite{Deutsch1996} in a certain regime of errors, and uses less spatial resources, but at the expense of a temporal overhead.
As done in the Refs.~\cite{Piparo2012, Razavi2009a}, we will divide the secret key rate by the number of memories needed per node. For the distillation strategies of the quantum repeater we consider a nested distillation scheme, i.e., where distillation after each swapping is performed. A special case will be distillation only before the first swapping, which might be experimentally more feasible. We thoroughly investigate the case where the number of distillation rounds in each nesting level is identical. Then, we lift this restriction and vary the number of distillation rounds individually after each swapping.
Additionally, we account for the classical communication time needed for acknowledging the success of entanglement swapping and entanglement distillation in the quantum repeater nodes. For this purpose we will derive a formula for the generation rate of the entangled pairs (repeater rate) including these classical communication times.

The paper is structured as follows: in Sec.~\ref{sec:QR} we review the concept of quantum repeaters, the relevant distillation protocols, and distillation strategies. In Section~\ref{sec:SKR} we present analytical formulas for the secret key rates. As the secret key rate is a product of the secret fraction and the repeater rate, we will derive the latter for the different distillation protocols. In Sec.~\ref{sec:keyrate} we analyze the quantum repeater in the context of quantum key distribution and present the optimal secret key rates. Here, the secret key rates are optimized with respect to the different distillation protocols and distillation strategies, the number of nesting levels, the number of distillation rounds, and the number of used memories. Furthermore, we investigate the impact of finite-efficiency detectors on the secret key rate. Then, we will fix the number of required memories and investigate the optimal setup.  In Sec.~\ref{sec:Impact} the influence of the classical communication time on the secret key rate is analyzed. We conclude in Sec.~\ref{sec:conclusion}.

\section{\label{sec:QR}Quantum repeater and distillation strategies}

\begin{figure}[htb]
 \centering
 \includegraphics[width=0.4\textwidth]{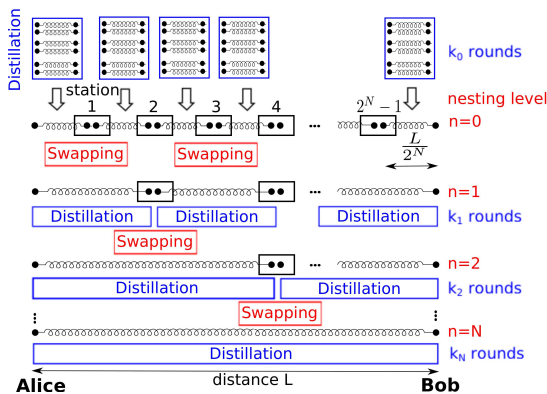}
 % qr_3.pdf: 601x487 pixel, 72dpi, 21.20x17.18 cm, bb=
 \caption{(Color online) A generic quantum repeater protocol with nested distillation (see text).}
 \label{fig:qr_3}
\end{figure}

In Fig.~\ref{fig:qr_3} we show a quantum repeater setup, whose concept was introduced in \cite{Briegel1998}. The goal is to establish an entangled pair between the two parties Alice and Bob over the distance $L$. For this reason, one divides the distance into segments of length $L_0=\frac{L}{2^N}$, where $N$ is the number of \textit{maximal nesting levels} for swapping. The segments are connected by repeater stations, which are able to perform Bell measurements and distillation. Due to entanglement swapping the fidelity
degrades, which we compensate by entanglement distillation. We define the fidelity of a state $\rho$ as its overlap with the Bell state $\left|\phi^+\right>=\frac{1}{\sqrt{2}}\left(\left|00\right>+\left|11\right>\right)$, i.e.,
\begin{equation} F(\rho):=\braccket{\phi^+}{\rho}{\phi^+},
\label{eq:fid}
\end{equation}
where $\ket{0}$ ($\ket{1}$) is, e.g., a horizontally (vertically) polarized photon.

In the following we will describe the distillation protocols that we want to compare. Our figure of merit for the comparison is the secret key rate, in contrast to Ref.~\cite{Dur1999}, where the final fidelity of the entangled state was investigated. As the secret key rate is not only a function of the fidelity, our conclusions for optimal distillation are different from Ref.~\cite{Dur1999}. In the following we will assume analogously to \cite{Dur1999} that the quantum gates are subjected to depolarizing noise with probability $(1-p_G)$ and with probability $p_G$ they are perfect\footnote{The formulas for the fidelity and the success probability considering this error parameter can be also found in \cite{Dur1999}. Different to \cite{Dur1999}, we do not assume any misalignment and the single-qubit operation is error free.}.

\subsection{The distillation protocols}

General distillation protocols consist of performing local operations on $n$ qubit pairs resulting in $m<n$ pairs with a higher fidelity than the initial pairs. Throughout this paper we will consider protocols that operate on two qubit pairs and lead to one qubit pair. Usually, local operations and a \textsc{cnot}-gate are applied. The sequence of these operations is specific for every protocol. Finally, both parties perform a measurement and depending on the outcome the resulting pair has a higher fidelity or is discarded. Thus, the protocols are probabilistic. In the following we briefly describe the protocols considered in this paper.

\subsubsection{Recurrence protocol: The \textit{Deutsch et al.} protocol}
\begin{figure}[!htpb]
 \centering
 \includegraphics[width=0.5\textwidth]{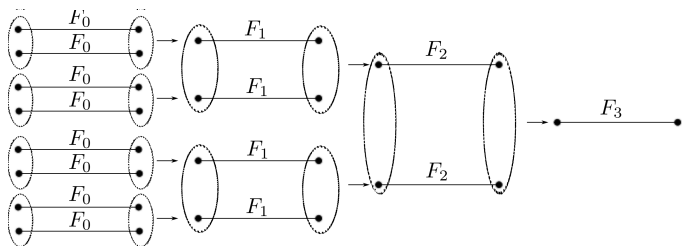}
 % ent_pump.pdf: 656x61 pixel, 72dpi, 23.14x2.15 cm, bb=0 0 656 61
 \caption{Recurrence protocol: \DP protocol (figure adapted from \cite{Dur1999}). The fidelity $F_k$ is the fidelity in the $k$-th distillation round.}
 \label{fig:recurrence}
\end{figure}

The \textit{Deutsch et al.}  protocol \cite{Deutsch1996}, sometimes called \textit{Oxford protocol}, works in a similar way as the distillation protocol introduced in \cite{Bennett1996a, Bennett1996}, but is more efficient. It can reach a higher fidelity in fewer distillation rounds, and therefore results in higher secret key rates. In general the protocol operates on Bell-diagonal states, i.e.,
\begin{equation}
\label{eq:bell}
\rho_{\rm Bell}=A \Pi_{\ket{\phi^+}}+B \Pi_{\ket{\phi^-}}+C \Pi_{\ket{\psi^+}}+D \Pi_{\ket{\psi^-}},
\end{equation}
with $A,B,C,D \geq0$, $A+B+C+D=1$, and  $\Pi_{\ket{\psi}}=\proj{\psi}$ being the projectors onto the four Bell states $\ket{\phi^\pm}=\frac{1}{\sqrt{2}}\left(\ket{00}\pm \ket{11}\right)$ and $\ket{\psi^\pm}=\frac{1}{\sqrt{2}}\left(\ket{01}\pm \ket{10}\right)$.
For each state of the form in Eq.~\eqref{eq:bell}, the first qubit belongs to Alice, the second to Bob. Both share two pairs of the state given in Eq.~\eqref{eq:bell}. Alice (Bob) applies a $\pi/2$ $(-\pi/2)$ rotation about the $X$-axis on her (his) two qubits, followed by a \textsc{cnot} operation on both sides. After that a bilocal measurement on one qubit in the computational basis is performed. 
 The values of the parameters $A,B,C,$ and $D$ as a function of the imperfections of the \textsc{cnot} and the fidelity $F$ can be found in \cite{Dur1998}.
The protocol works in a recursive way, i.e., it uses two copies of the same fidelity for the next distillation step; therefore it is called \textit{recurrence protocol} (see Fig.~\ref{fig:recurrence}).

\subsubsection{Entanglement pumping: The \textit{D\"{u}r et al.} protocol}
\begin{figure}[ht]
 \centering
 \includegraphics[width=0.5\textwidth]{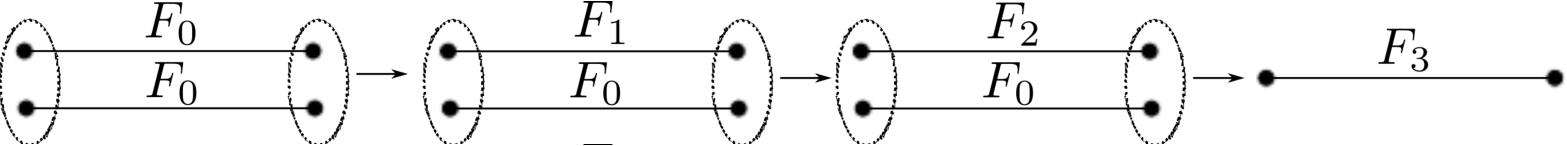}
 % ent_pump.pdf: 656x61 pixel, 72dpi, 23.14x2.15 cm, bb=0 0 656 61
 \caption{Entanglement pumping: \textit{D\"{u}r et al.} protocol (figure adapted from \cite{Dur1999}).}
 \label{fig:ent_pump}
\end{figure}

This protocol introduced in \cite{Dur1999}, sometimes also called \textit{Innsbruck protocol}, uses the \textit{Deutsch et al.} protocol, but the two input states do not need to have the same fidelity. Here, distillation is performed with an auxiliary pair having always the same initial fidelity $F_0$, see Fig.~\ref{fig:ent_pump}, hence the name \textit{entanglement pumping}. We see that different to the \DP protocol the number of required memories does not depend on the number of rounds of distillation, but it is linear in the number of nesting levels (see Sec.~\ref{subsec:mem}).

Throughout the paper we will assume that we only start with entanglement swapping and entanglement distillation when both pairs are present.  
\subsection{\label{sec:strategies}Distillation strategies for the quantum repeater}
\begin{figure}[htb]
 \centering
 \includegraphics[width=0.4\textwidth]{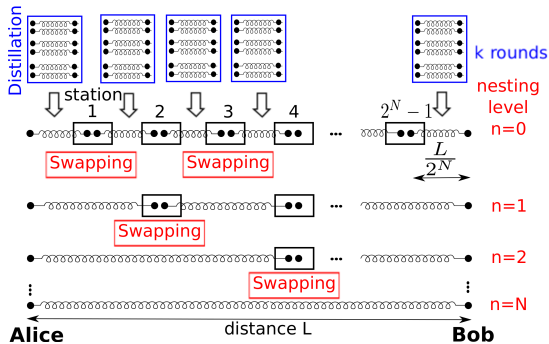}
 % qr_3_distBegin.pdf: 582x368 pixel, 72dpi, 20.53x12.98 cm, bb=
 \caption{(Color online) \textit{Distillation strategy $\beta$}: distillation only in the beginning.}
 \label{fig:qr_3_distBegin}
\end{figure}

The protocols described in the previous section can be inserted into the quantum repeater protocol in different ways. In the following we want to compare two different specific distillation strategies. For this purpose we define the distillation vector 
\begin{equation}
\label{eq:kvec}
\vec{k}=(k_0,...,k_{N}) 
\end{equation}
for the distillation rounds, where each component with index $n$ gives the number of distillation rounds in the $n$-th nesting level (see Fig.~\ref{fig:qr_3}). Throughout the paper \textit{distillation strategy $\alpha$} denotes a strategy with the same number of distillation rounds in each nesting level, hence the distillation vector is $\vec{k}^\alpha=(k,...,k)$. A strategy which might be less demanding for experimental realizations\footnote{When only swapping is performed, one can collect the outcomes of the Bell-measurements and later apply bit-flips on the classical data resulting from the QKD-measurement on the final state (see also \cite{Abruzzo2012}). For the case of distillation after swapping the single-qubit rotations have to be applied, thus the number of quantum operations is increased.}, is the \textit{distillation strategy $\beta$} (see Fig.~\ref{fig:qr_3_distBegin}), where we only distill at the beginning. The distillation vector is thus $\vec{k}^\beta=(k,0,...,0)$. In Sec.~\ref{sec:distStr} we will use general distillation vectors. This strategy will be called \textit{distillation strategy $\gamma$}.

%\section{\label{sec:keyrate} Optimal secret key rates: Comparing different distillation protocols and strategies}

\section{\label{sec:SKR}Secret key rates and the quantum repeater}
In the previous section we have described the generation of entangled pairs over a distance $L$ between the parties Alice and Bob using the quantum repeater protocol. For performing quantum key distribution (QKD), they measure each of their particles in some measurement basis. In this paper we consider the six-state protocol \cite{Bruß1998,Bechmann-Pasquinucci1999}; the BB84-protocol \cite{Bennett1984} leads to similar secret key rates. The former works as follows: for each qubit pair Alice and Bob each perform measurements in the $X$-, $Y$- and $Z$-direction. After the measurement the used basis is announced (\textit{sifting phase}). Only those measurement results where their measurement bases coincided will be utilized in the further analysis. %for the key and a part of it will be utilized for estimating the error rate, i.e. the ratio of discordant bits and the total number of bits. 
We adopt here the asymmetric protocol \cite{Lo2005} which uses different probabilities for the choice of the measurement direction. In this protocol the sifting parameter, i.e., the fraction of sifted bits, is one in the asymptotic limit, which we also assume here.
The quantum bit error rate (QBER), i.e., the fraction of discordant bits, bounds the eavesdropping attempt: if it is above a certain threshold the protocol is aborted. The quantity we are interested in is the \textit{secret key rate} $K$ per memory per second, which is the product of the \textit{repeater rate} $R_{\rm Rep}$ and the \textit{secret fraction} $r_{\infty}$ (see, e.g., \cite{Scarani2009} for a review) divided by the number of memories:
\begin{equation}
\label{eq:KMem}
 K^i=R_{\rm Rep}^i r_\infty /M^i,
\end{equation}
with the superscript $i$ being either D (\DP protocol) or D\"{u}r (\DurP protocol). 

In the following sections we will describe or derive each component of the secret key rate given in Eq.~\eqref{eq:KMem}.

%As mentioned above, the secret key rate $K$ consists of the repeater rate $R_{\rm Rep}$ and the secret fraction $r_{\infty}$. 
\subsection{The secret fraction}
The secret fraction is the ratio of secret bits and the measured bits in the asymptotic limit, thus denoted by $r_{\infty}$. It is given by the so-called Devetak-Winter bound \cite{Devetak2005} and can be expressed in terms of the error rates appearing in the six-state protocol \cite[Appendix A]{Scarani2009}:
\begin{eqnarray}
\label{eq:6S}
r_\infty&=&1-e_{Z}h\left(\frac{1+(e_{X}-e_{Y})/e_{Z}}{2}\right)\nonumber\\
&&-(1-e_{Z})h\left(\frac{1-(e_X+e_Y+e_Z)/2}{1-e_Z}\right)\nonumber\\
&&-h(e_{Z}),
\end{eqnarray}
with $h(p)=-p \log_2 p-(1-p) \log_2 (1-p)$ being the binary Shannon entropy and $e_X$, $e_Y$ and $e_Z$  being the error rates in the $X$-, $Y$-, and $Z$-basis, respectively. For a detailed analysis of the topic of quantum key distribution in connection to quantum repeaters we refer to \cite{Abruzzo2012}. Note that the secret fraction depends on the following parameters: the initial fidelity $F_0$, the gate quality $p_G$, the maximal nesting level $N$, the distillation vector $\vec{k}$, and the distillation protocol.

\subsection{\label{sec:rep}The repeater rate, including classical communication times}
By the repeater rate $R_{\rm Rep}$ we denote the average number of long-distance entangled pairs generated by the quantum repeater per second. Considering a setup which connects only the neighboring pairs (so-called parallelization), several formulas for different physical realizations of a quantum repeater were derived: Ref.~\cite{Bernardes2011} treats the repeater rate for deterministic swapping and probabilistic distillation before the first swapping, Ref.~\cite{Sangouard2011} deduces the rate for probabilistic swapping without distillation and in \cite{Abruzzo2012} the formula from the latter reference was modified to allow distillation before the first swapping. These expressions have in common that they do not consider the classical communication times needed to acknowledge the success of entanglement swapping and entanglement distillation. In the following we will derive a repeater rate for probabilistic swapping and probabilistic distillation including these communication times. Our derivation is inspired by the recurrence formula developed for quantum repeaters based on nitrogen-vacancy (NV) centers in diamond \cite{Childress2005}. In Sec.~\ref{sec:Impact} we show how the secret key rate changes, when we omit the classical communication times needed for entanglement swapping and entanglement distillation. We will always assume that the entanglement distribution requires classical communication. 

\subsubsection{The \DP protocol}
We define the repeater rate to be the reciprocal value of the time $\tau^\textrm{D}(\kvec,N)$ needed to establish an entangled pair over the distance $L$, with $N$ being the maximal nesting level and the distillation vector $\vec{k}=(k_0,...,k_N)$, i.e.,
\begin{equation}
 \label{eq:rawDCC}
R_{\rm Rep}^\textrm{D}:=\frac{1}{T_0 \tau^\textrm{D}(\kvec,N)}.
\end{equation}
Here, the superscript D refers to the \DP protocol. Note that the time $\tau^\textrm{D}(\kvec,N)$ is given in units of the fundamental time $T_0:=\frac{L_0}{c}$, with $c=2\cdot 10^5$ km/s the speed of light in the optical fiber and $L_0:=\frac{L}{2^N}$ the fundamental length, i.e., the distance between the repeater stations. The symbol $\tau^\textrm{D}(k_N,N)$ with only one vector component $k_N$ as first argument denotes the time needed in nesting level $N$ for $k_N$ distillation rounds. In the following we present a recurrence formula for $\tau^\textrm{D}(k_N,N)$ given by:
\begin{subequations}
\label{allequations}
\begin{eqnarray}
\tau^\textrm{D}(k_0=0,N=0)&=&\frac{2}{P_0}\label{equationc},\\ 
\tau^\textrm{D}(k_N=0,N>0)&=&\frac{1}{P_{ES}(N)}\left[\frac{3}{2}\tau^\textrm{D}(k_{N-1},N-1)\right.\nonumber\\
&&\left.+2^{N-1}\right],\label{equationb}\\
\tau^\textrm{D}(k_N>0,N)&=&\frac{1}{P_D^\textrm{D}(k_N,N)}\left[\frac{3}{2}\tau^\textrm{D}(k_N-1,N)\right.\nonumber\\
&&\left.+2^N\right],\label{equationa}
\end{eqnarray} 
\end{subequations} 
with $P_{ES}(N)$ being the success probability of entanglement swapping in the $N$-th nesting level and $P_D^\textrm{D}(i,N)$ being the probability of success for entanglement distillation using the \DP protocol in the $i$-th distillation round in the $N$-th nesting level. Here $P_0$ is the probability to generate an entangled photon pair over a distance $L_0$ and is given by $P_0=10^{-\alpha L_0/10}$, with $\alpha=0.17$ dB/km being the attenuation coefficient.
To explain the recurrence formula in Eq.~\eqref{allequations}, we start from the first line [Eq.~\eqref{equationc}]. There, we assume that the source is placed at one side and the photon is distributed over the distance $L_0$ leading to a distribution time of $T_0$. The acknowledgment of the arrival of the photons needs at least the same time, so we have in total $2T_0$ (see \cite{Abruzzo2012} for further details and other schemes of entanglement distribution). We divide by the probability $P_0$ to generate this entangled photon pair as on average we have to perform this process $\frac{1}{P_0}$ times (see, e.g., \cite{Sangouard2011} for an explicit calculation of this waiting time). The next line [Eq.~\eqref{equationb}] gives the time for the $N$-th nesting level before starting with distillation, i.e., it is the time directly after entanglement swapping. The formula consists of two parts: the generation time for the pairs needed to begin the swapping $\left[\frac{3}{2}\tau^\textrm{D}(k_{N-1},N-1)T_0\right]$ (see, e.g., \cite[Appendix A]{Sangouard2011} for an explanation of the factor $\frac{3}{2}$) and the time to acknowledge the success of the swapping, i.e., $2^{N-1}T_0$; both divided by the probability of 
successful swapping in the $N$-th nesting level $\frac{1}{P_{ES}
(N)}$. Note that the factor $\frac{3}{2}$ is an approximation for small probabilities. The first part $\left[\frac{3}{2}\tau^\textrm{D}(k_{N-1},N-1)T_0\right]$ corresponds to the average time to generate two pairs after $k_{N-1}$ rounds of distillation in the $(N-1)$-th nesting level. The last line [Eq.~\eqref{equationa}] concludes the recurrence formula: we need the time $\frac{3}{2}\tau^\textrm{D}(k_N-1,N)T_0$ to generate two pairs for the $k_N$-th round of distillation. As distillation is performed over the distance $\frac{L}{2^N}$ the acknowledgment time is $2^N T_0$. Both terms are divided by the probability of success for entanglement distillation $\left[P_D^\textrm{D}(k_N,N)\right]$.

We present the analytic solution of the recurrence formula in Eq.~\eqref{allequations} in the Appendix~\ref{subsec:rawkey}, Eq.~\eqref{sol:DRepCC}.

\subsubsection{The \DurP protocol}
The repeater rate for the \DurP protocol differs from the repeater rate for the \DP protocol, as the entanglement distillation process works in a sequential way, i.e., the auxiliary pair for each distillation round is always the same (see Fig.~\ref{fig:ent_pump}). As the swapping process is the same in both distillation protocols, Eqs.~\eqref{eq:reca} and \eqref{eq:recb} are analogous to Eqs.~\eqref{equationc} and \eqref{equationb}: 
%  This explains why the recurrence formulas for the time $\tau^\textrm{D\"{u}r}(k_N,N)$ needed for \DurP are almost equal to Eq.~\eqref{allequations}: 
\begin{subequations}
\label{eq:rec}
\begin{eqnarray}
\tauDur(k_0=0,N=0)&=&\frac{2}{P_0},\label{eq:reca}\\
\tauDur(k_N=0,N>0)&=&\frac{1}{P_{ES}(N)}\left[\frac{3}{2}\tauDur(k_{N-1},N-1)\right.\nonumber\\
&&\left.+2^{N-1}\right]\label{eq:recb}\\
\tauDur(k_N>0,N)&=& \frac{1}{P_D^\textrm{D\"{u}r}(k_N,N)}\left[\tauDur(k_N-1,N)\right.\nonumber\\
&&\left.+\tauDur(0,N)+2^N\right]\label{eq:recd}.
\end{eqnarray}
\end{subequations}

The third line [Eq.~\eqref{eq:recd}] differs from Eq.~\eqref{equationa}. Equation~\eqref{eq:recd} represents the time needed to distill a pair in the $k_N$-th round in the $N$-th nesting level. In the entanglement pumping protocol, we start to produce the elementary pair $\rho(k_N=0,N)$ for distillation when the pair to be distilled, $\rho(k_N-1,N)$, is present. Thus, we have to add the time for generating the elementary pair $\tauDur(0,N)T_0$ to the time for the pair to be distilled, $\tauDur(k_N-1,N)T_0$.
% we take the pair that was distilled in the previous round, i.e. the time $\tauDur(k_N-1,N)T_0$ and the elementary pair, which is not distilled in the $N$-th nesting level, i.e.\ the time is $\tauDur(0,N)T_0$.
The repeater rate for the \DurP is then given by
\begin{equation}
 \label{eq:rawDurCC}
R_{\rm Rep}^{\text{D\"{u}r}}:=\frac{1}{T_0 \tauDur(\kvec, N)}.
\end{equation}

We give an analytic solution of the recurrence formula in the Appendix, Eq.~\eqref{eq:taukN}.

\subsection{\label{subsec:mem}Number of memories}
In this section we describe the needed number of memories at each half of the repeater station (see the black dots in Fig.~\ref{fig:qr_3}). The vector $\vec{k}$ consists of the number $k_n$ of distillation rounds in the $n$-th nesting level, see Eq.~\eqref{eq:kvec}. The number of memories needed at half a node for the \DP protocol is 
\begin{equation}
\label{eq:MD}
M^{\rm D}=2^{\sum_n k_n},
\end{equation}
because in each nesting level the number of memories needs to be increased by a factor of $2^{k_n}$, as the distillation for all nesting levels is done in parallel. The superscript D denotes the \DP protocol.

The \DurP protocol works in a sequential way, so the number of memories is
\begin{equation}
\label{eq:MDur}
M^{\text{D\"{u}r}}=N+2-|\{k_i:k_i=0\}|,
\end{equation}
where the set $|\{k_i:k_i=0\}|$ is the number of elements in $\kvec$ that are zero.
Equation~\eqref{eq:MDur} for strategy $\alpha$, i.e., $\kvec=(k,k,...,k)$, can be explained as follows: For nesting level $N=0$, at most two memories are needed for the distillation process (see Fig.~\ref{fig:ent_pump}). The resulting pair $\rho(k_0,N=0)$ at distance $L_0$ after $k_0$ distillation rounds is stored in one memory, and the other one is emptied. After swapping two neighboring pairs we have the pair $\rho(0,N=1)$ at the distance $2L_0$. For starting the distillation process in this nesting level ($N=1$), one needs another pair $\rho(0,N=1)$, which is generated by the same procedure as above, so two additional memories are needed. In total one needs three memories for $N=1$.
% After performing distillation one stores the particle in one memory and performs swapping. For the distillation process one has to use another memory, so one always has one memory more than the nesting level. 
For strategy $\beta$, i.e., $\kvec
=(k,0,...,0)$, one just needs two memories where we store the state during the gate operation.

\section{\label{sec:keyrate} Optimal secret key rates: Comparing different distillation protocols and strategies}

\subsection{\label{sec:mem}Comparison of key rates (strategy $\alpha$ vs $\beta$)}

%  We start with two states, each having a depolarized form 
% \begin{equation}
% \label{eq:rhoDep}
% \rho_{Dep}=F \Pi_{\ket{\phi^+}}+\frac{1-F}{3}\left(\Pi_{\ket{\phi^-}}+\Pi_{\ket{\psi^+}}+\Pi_{\ket{\psi^-}}\right),
% \end{equation}

We investigate how the \DP and the \DurP protocol perform under gate errors where we use the secret key rates as a figure of merit.

In the following we calculate the secret key rate divided by the number of needed memories [see Eq.~\eqref{eq:KMem}]. The division by the number of memories allows for a fair comparison when considering the resources. For a fixed set of parameters $F_0$ (initial fidelity) and $p_G$ (gate quality) we aim at finding the optimal distillation protocol, the optimal number of distillation rounds, the optimal number of nesting levels, the best distillation strategy, and the minimal number of memories. Note that in the ideal case, i.e., for perfect detectors, we assume the entanglement swapping to be deterministic, i.e., $P_{ES}(N)=1$.

We will consider two error models for the input states: on one hand depolarized states, and on the other hand so-called binary states. The latter states are interesting, as they can be produced by the hybrid quantum repeater \cite{VanLoock2006, Ladd2006}. Additionally, in \cite{Dur1999} it was mentioned that the binary state given in Eq.~\eqref{eq:bin} below has the optimal shape for the \DurP protocol. 

\subsubsection{\label{subsec:dep}Input states: Depolarized states}

In this section we want to investigate the optimal secret key rates [Eq.~\eqref{eq:KMem}] when we start with depolarized states, i.e.,
\begin{equation}
\label{eq:rhoDep}
\rho_{Dep}=F \Pi_{\ket{\phi^+}}+\frac{1-F}{3}\left(\Pi_{\ket{\phi^-}}+\Pi_{\ket{\psi^+}}+\Pi_{\ket{\psi^-}}\right).
\end{equation}

\begin{figure*}[!htpb]
\centering
\includegraphics[width=0.6\textwidth, clip, angle=270]{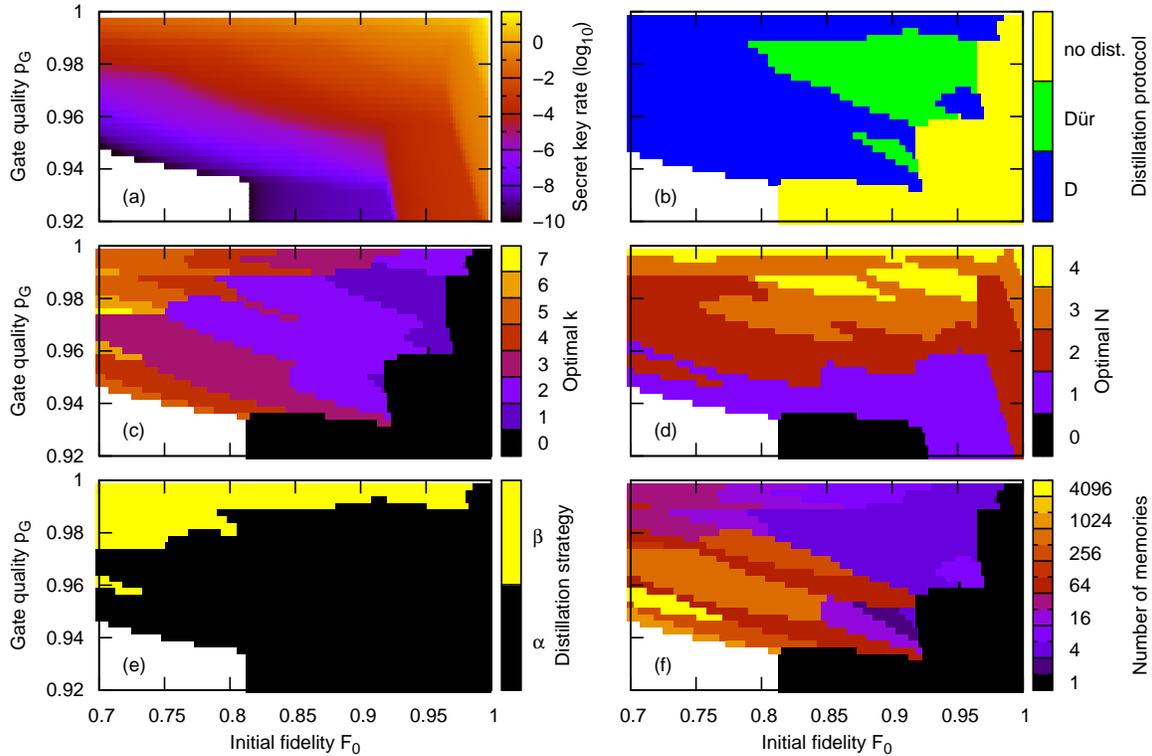}
 \caption{(Color online) (a) Optimal secret key rate per memory per second [Eq.~\eqref{eq:KMem}] for the distance $L=600$ km. The smallest secret key rate still depicted is chosen to be $10^{-10}$ secret bits per second per memory. In the white region an extraction of a non-zero secret key rate is not possible. The parameters for the optimal secret key rate per memory per second are: (b) Distillation protocols: \DP protocol (blue), \DurP protocol (green), and no distillation (yellow). (c) Number of rounds of distillation $k$ (for the optimal distillation strategy). (d) Number of nesting levels $N$. (e) Distillation strategies: Strategy $\alpha$ (nested distillation) and strategy $\beta$ (distillation only before the first entanglement swapping). (f) Number of used memories per repeater node.}
 \label{fig:multiplot}
\end{figure*}

% \begin{figure}[!htpb]
% \centering
% \includegraphics[width=0.275\textwidth, clip, angle=270]{img/OptMemScrScan_key.ps}
%  \caption{(Color online) Optimal secret key rate per memory per second [Eq.~\eqref{eq:KMem}, $L=600$ km]. The smallest secret key rate still depicted is chosen to be $10^{-10}$ secret bits per second per memory.}
%  \label{fig:OptMemScrScan_key}
% \end{figure}

Optimization of the distillation protocols (\DP or \DurP), the number of nesting levels $N$, the number of distillation rounds $k$, and the distillation strategy ($\alpha$ or $\beta$), leads to the secret key rates depicted in Fig.~\ref{fig:multiplot}~(a). We point out that we find the global maximum as we calculate $K^i$ for all possible combinations of parameters for the length $L$ and then choose the maximal value. The parameters leading to the optimal secret key rates of Fig.~\ref{fig:multiplot} (a) are shown in Figs.~\ref{fig:multiplot}~(b)-(f).
The optimal distillation protocol is shown in  Fig.~\ref{fig:multiplot}~(b). It is difficult to find an intuitive explanation why in certain regimes either the \DP or the \DurP protocol is optimal; there are many different effects such as the repeater rates [see Eqs.~\eqref{allequations} and \eqref{eq:rec}], the number of memories and the resulting state.
Figure~\ref{fig:multiplot}~(c) shows the optimal number of distillation rounds (for the optimal distillation strategy) that lead to the secret key rate per memory per second of Fig.~\ref{fig:multiplot}~(a). We find that for a wide range of parameters it is enough to have $k\leq 3$ distillation rounds.
The role of the optimal number of nesting levels is treated in Fig.~\ref{fig:multiplot}~(d). We find that with increasing gate quality and initial fidelity more nesting levels are optimal. 
In figure~\ref{fig:multiplot}~(e) the optimal of the two distillation strategies ($\alpha$) or ($\beta$) is shown: for good gates and low fidelities it is better to only distill in the beginning, which would be experimentally less demanding. We emphasize that in this regime of parameters distillation in later nesting levels degrades the secret key rate.
From the previous plots, we calculate in Fig.~\ref{fig:multiplot}~(f) the minimal number of memories needed to obtain the secret key rate in Fig.~\ref{fig:multiplot}~(a).

\begin{figure}[!htbp] % Example of including images
\begin{center}
\includegraphics[width=0.32\textwidth, clip, angle=270]{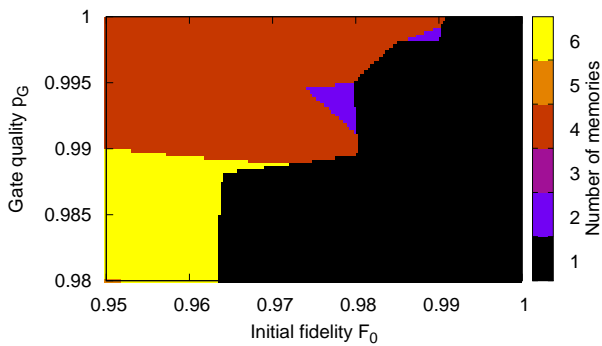}
\end{center} 
\caption{(Color online) Expanded region from Fig.~\ref{fig:multiplot}~(f): Number of memories that lead to the optimal secret key rate per second per memory [see Eq.~\eqref{eq:KMem}, $L=600$ km].}
\label{fig:OptMemScrScan_memoriesZoom}
\end{figure}

Figure~\ref{fig:OptMemScrScan_memoriesZoom} provides a zoom of Fig.~\ref{fig:multiplot}~(f) into the region, where the secret key rate is in the order of bits per second. In the black region no distillation is optimal, therefore we only need one memory. For the number of memories $M=2$ and $M=4$, the optimal protocol is the \DP protocol, whereas for $M=6$ the \DurP protocol becomes favorable. We see from Eq.~\eqref{eq:MD} that in a single setup  the number of memories is restricted to a power of two for the \DP protocol. If we want to use, e.g., $M=6$ memories and the \DP protocol, we have to employ setups in parallel. We will treat this subject in Sec.~\ref{subsec:fixedMem}.

\subsubsection{Input states: Binary states}
We will now consider binary states, i.e., states of the form 
\begin{equation}
\rho_{Bin}=F\proj{\phi^+}+(1-F)\proj{\phi^-}.
\label{eq:bin}
\end{equation}
We performed a complete analysis of this case, in analogy to subsection~\ref{subsec:dep}. The results of our investigation can be summarized as follows:
\begin{itemize}
 \item Different to the setup where we start with depolarized states, it is possible to extract a non-zero secret key rate per memory per second for the whole range of parameters considered here, i.e., for $0.7\leq F_0\leq1$ and $0.92\leq p_G\leq 1$. The largest value of the secret key rate per memory per second using binary states is in the same order of magnitude as for depolarized states.
\item The region where the \DurP protocol is optimal extends to lower initial fidelities, compared to Fig.~\ref{fig:multiplot}~(b) and the largest value for the optimal rounds of distillation is $k=3$. Also the region where no distillation is optimal increases.
\item Due to the small optimal $k$, the maximal number of memories decreases. 
\end{itemize}
One would recommend the use of binary states when $p_G\leq0.97$ and $F_0\leq0.8$, as then the number of used memories is smaller than for depolarized states and additionally the secret key rate per memory per second is non-zero.
%  But one must be careful as the probability to generate binary states in the case of the hybrid quantum repeater is a function of the fidelity and as a consequence the secret key rate per second is not a monotonic function of the fidelity (see \cite{Abruzzo2012} for additional information). Thus, additional optimization should be employed.

\subsection{\label{sec:det}The influence of the detector efficiency}
In this section we want to investigate the impact of finite-efficiency detectors on the secret key rate. The detector efficiency is given by the parameter $\etaD$, with $0\leq\etaD\leq1$, where $\etaD=1$ corresponds to perfect detectors.
%  We consider a detector efficiency of $\eta_d=90\%$, which can be reached for a wavelength of $1556$ nm with superconducting transition-edge sensors (TESs) (see \cite{Eisaman2011} and references therein) at the cost of very low temperatures ($0.1$ K) and $\eta_d=10\%$ achieved by InGaAs photodetectors operating at a wavelength of $\lambda=1550$ nm (see \cite{Eisaman2011} for recent review on single-photon sources and detectors).
 For implementing the detector efficiency in our formulas we have to replace the probability of successful distillation $P_D(k,n)$ and the probability of successful swapping in the $n$-th nesting level $P_{ES}(n)$ in the equations for the repeater rate [Eqs.~\eqref{eq:rawDCC} and~\eqref{eq:rawDurCC}] by
\begin{subequations}
\label{eq:deteff}
\begin{eqnarray}
 P_D(k,n)&\rightarrow& \eta_d^2 P_D(k,n)\\
 P_{ES}(n)&\rightarrow& \eta_d^2  P_{ES}(n),
\end{eqnarray}
\end{subequations}
because the Bell-measurement requires a two-fold detector click. Additionally, we have to multiply the secret key rate [Eq.~\eqref{eq:KMem}] by a factor of $\eta_d^2$ which accounts for the final quantum key distribution measurement.

The only contribution of the detector efficiency in the secret key rate is in the repeater rate. For simplicity we will consider the repeater rate without classical communication for entanglement swapping and entanglement distillation [see Eq.~\eqref{eq:raw} and ~\eqref{eq:rawDur} in Sec.~\ref{sec:Impact}]. After replacing the probabilities in the repeater rates by Eq.~\eqref{eq:deteff}, the repeater rate scales with $\etaD^{2(N+\sum_n k_n)}$.

\begin{figure}[!htpb]
\includegraphics[width=0.32\textwidth, clip, angle=270]{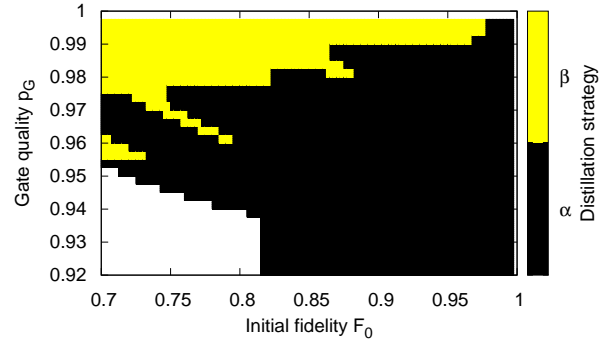}
\caption{(Color online) Distillation strategies with imperfect detectors: Strategy $\alpha$ (nested distillation strategy) and strategy $\beta$ (distillation only before the first entanglement swapping) that lead to the optimal secret key rate per memory per second [Eq.~\eqref{eq:KMem}, $L=600$ km and $\eta_d=0.9$].}
\label{fig:OptMemScrScanDet90_distStr}
\end{figure}

When analyzing different detector efficiencies we made the following observations:
\begin{itemize}
 \item with decreasing $\eta_d$, the region where no distillation is optimal increases  such that for $\eta_d=0.1$, it is optimal to not perform distillation for almost all parameters,
\item with decreasing $\eta_d$, the optimal number of nesting levels also decreases,
\item with decreasing $\eta_d$, the region where the distillation strategy $\beta$ (distillation only in the beginning) is optimal increases (see Fig.~\ref{fig:OptMemScrScanDet90_distStr}).
\end{itemize}
%We can estimate from the repeater rates given in App.~\ref{subsec:rawkey} that the secret key rate scales with $\eta_d^{2N+2\sum_{n=0}^{N} k_n+2}$. 
Figure~\ref{fig:OptMemScrScanDet90_distStr} shows the optimal distillation strategies for the secret key rate per memory per second with a detector efficiency of $\eta_d=0.9$. This can be compared to Fig.~\ref{fig:multiplot}~(e), where the detectors are perfect, i.e., $\eta_d=1.$ We see that for low initial fidelities the region where the distillation strategy $\beta$ is optimal increases.

\subsection{More general strategies}
\subsubsection{\label{sec:distStr} Distillation strategy $\gamma$}

As mentioned in Sec.~\ref{sec:strategies} we now lift the restriction that the number of distillation rounds in each nesting level is the same. For this purpose we fix the parameters for the initial fidelity $F_0$ and the gate quality $p_G$ and vary the number of nesting levels and the number of distillation rounds in each nesting level. A result for the parameters $F_0=0.9$ and $p_G=0.96$, is shown in Table~\ref{Tab2}.
\begin{table}[h]%[H] add [H] placement to break table across pages
\caption{\label{Tab2}Optimal secret key rate per memory per second [Eq.~\eqref{eq:KMem}] and corresponding distillation vector $\vec{k}$ [Eq.~\eqref{eq:kvec}] for the different distillation protocols, $F_0=0.9$, and $p_G=0.96$.}
\begin{ruledtabular}
\begin{tabular}{ccccc}
 &\multicolumn{2}{c}{\DurP protocol}&\multicolumn{2}{c}{\DP protocol} \\
N& $K$ & $\vec{k}$ &  $K$ & $\vec{k}$ \\\hline
0 & $3.92\cdot10^{-9}$ & $(0)$&$3.92\cdot10^{-9}$ & $(0)$ \\ 
1 & $2.11\cdot10^{-5}$ & $(0,2)$&$2.63\cdot10^{-5}$ & $(0,1)$ \\ 
2 &$1.09\cdot10^{-4}$ & $(2,3,2)$&$3.03\cdot10^{-4}$ & $(0,3,1)$ \\ 
3 & $2.66\cdot10^{-6}$ & $(3,4,5,5)$&$1.51\cdot10^{-4}$ & $(0,3,3,1)$ \\ 
4 & $0$ & $0$ &$1.37\cdot10^{-5}$ & $(0,3,3,3,1)$
\end{tabular}
\end{ruledtabular}
\end{table}
% \begin{table}[!htpb]
%  \begin{center}
% % use packages: array
% \begin{tabular}{l|l|l}
% N & $K_{\rm Sec}$ & $\vec{k}$ \\ 
% \hline
% 0 & $3.92\cdot10^{-9}$ & (0) \\ 
% 1 & $8.9\cdot10^{-5}$ & (0,2) \\ 
% 2 & $6.2\cdot10^{-3}$ & (0,3,2) \\ 
% 3 & $0.02$ & (0,3,3,2) \\ 
% 4 & $0.014$ & (0,3,3,3,1)
%  \end{tabular}
%  \end{center}
% \caption{\label{Tab2}Optimal secret key rate per second (Eq.~\eqref{eq:KSec}) and corresponding distillation vector $\vec{k}$ (Eq.~\eqref{eq:kvec}) for $F_0=0.9$, $p_G=0.96$ and the \DP protocol.}
% \end{table}
There, we report the optimal distillation vector $\vec{k}$, see Eq.~\eqref{eq:kvec}, for the number of nesting levels up to $N=4$, and the corresponding secret key rate per memory per second. We found the optimal $\vec{k}$ by calculating the key rate for all possible $\vec{k}$.
For the given parameters, distillation only in the beginning does not help. Comparing the values that we achieved in Sec.~\ref{sec:mem}, i.e., only considering strategy $\alpha$ [distillation vector $\vec{k}=\left(k,k,...,k\right)$] or $\beta$ [$\vec{k}=\left(k,0,...,0\right)$], the optimal secret key rate for the given set of parameters was $0.99\cdot 10^{-4}$ with $N=2$, $\vec{k}=(2,2,2)$ for the \DurP protocol. Here, the best secret key rate is $3.03\cdot 10^{-4}$ for $N=2,\; \kvec=(0,3,1)$ and the \DP protocol. Thus, the secret key rate is in the same of order of magnitude, but can be improved by a factor of 3.
\begin{table}[h]%[H] add [H] placement to break table across pages
\caption{\label{Tab3}Optimal secret key rate per memory per second [Eq.~\eqref{eq:KMem}] and corresponding distillation vector $\vec{k}$ [Eq.~\eqref{eq:kvec}] for the different distillation protocols, $F_0=0.97$, and $p_G=0.99$.}
\begin{ruledtabular}
\begin{tabular}{ccccc}
 &\multicolumn{2}{c}{\DurP protocol}&\multicolumn{2}{c}{\DP protocol} \\
N& $K$ & $\vec{k}$ &  $K$ & $\vec{k}$ \\\hline
0 & $7.97\cdot10^{-9}$ & $(0)$&$7.97\cdot10^{-9}$ & $(0)$ \\ 
1 & $9.64\cdot10^{-4}$ & $(0,0)$&$9.64\cdot10^{-4}$ & $(0,0)$ \\ 
2 & $0.19$ & $(0,0,0)$&$0.19$ & $(0,0,0)$ \\ 
3 & $0.57$ & $(0,0,2,0)$&$0.73$ & $(0,2,0,0)$ \\ 
4 & $0.96$ & $(0,1,1,1,0)$ &$0.88$ & $(0,1,1,1,0)$\\
5&$0.62$ & $(0,1,1,2,0,0)$&$0.54$&$(0,0,2,1,0,0)$\\
6 &$0.34$ & $(0,1,1,1,1,1,0)$&$0.2$&$(0,1,1,1,1,1,0)$
\end{tabular}
\end{ruledtabular}
\end{table}
Table~\ref{Tab3} gives results for the parameters $F_0=0.97$ and $p_G=0.99$. The parameters that lead to the optimal secret key rate per memory per second of $K=0.32$ in Sec.~\ref{sec:mem} are for the nesting level $N=3$, distillation strategy $\beta$, $\kvec=(2,0,0,0)$ using the \DP protocol. We see in this example that by allowing general distillation strategies, the optimal secret key rate can be increased by increasing the nesting level. In this example, different to above, the \DurP protocol remains optimal.

%We checked several parameters and came to the conclusion that a more general distillation strategy leads to an improvement not more than one order of magnitude.

Due to the computational complexity, we only calculated the general distillation strategies for two specific set of parameters (see Tables~\ref{Tab2} and \ref{Tab3}). As the quantum repeater exhibits a self-similar structure, dynamical programming was used in \cite{Jiang2007} in order to optimize the average time to create an entangled pair for a given final fidelity and distance. The results and methods of \cite{Jiang2007} cannot be used for a global optimization, as we have found counterexamples, where the distillation vector consist of different numbers in each nesting level (see, e.g., Table~\ref{Tab2} for the \DurP protocol and Table~\ref{Tab3}).

We see that it is not trivial to make general statements about the optimal number of rounds of distillation, regarding the secret key rate. For implementations, one has to determine the parameters of the experiment, i.e., $F_0$ and $p_G$, and then to optimize the secret key rate for any specific set of parameters.
\subsubsection{\label{subsec:fixedMem}Optimal strategies for a fixed number of memories allowing parallel setups}

We have mentioned in Sec.~\ref{subsec:mem} that in the following we want to fix the number of memories, and find which setup is optimal. As the memories in the \DP protocol are restricted to a power of two (see Sec.~\ref{subsec:mem}), we allow also setups working in parallel.

For calculating the optimal strategy for a fixed number of memories $M$, we solve the following equation to get all possible setups:
\begin{eqnarray}
\sum_{m=1}^{M} s_m m&=M
\label{eq:Mtotal}
 \end{eqnarray}
for $s_m\in\mathbb N$ and $\left\lfloor\frac{M}{m}\right\rfloor\geq s_m\geq 0$. The number $s_m$ denotes how many setups using $m$ memories work in parallel.
We then proceed by calculating for each setup the optimal secret key rate per second, i.e., $m K_m=r_\infty R_{\rm rep}$. The index $m$ for the secret key rate $K$ means that we restrict to distillation vectors and nesting levels that solve Eqs.~\eqref{eq:MD} and \eqref{eq:MDur} for $m$ memories. The optimal vector $\vec{s}=\left(s_1,...,s_{M}\right)$, a solution of Eq.~\eqref{eq:Mtotal}, is found by maximizing the value $\sum_m s_m m K_m$. The secret key rate of the total setup with a fixed number of memories $M$ is thus given by
\begin{equation}
K=\frac{\sum_m s_m mK_m}{M},
\label{eq:KMemtotal}
\end{equation}
with $\sum_m s_m m=M$.
We will also compare this result to a configuration of one setup with distillation vector $\vec{k}$ [see Eq.~\eqref{eq:kvec}], if possible. For the parameters $F_0=0.97$ and $p_G=0.99$ we calculated the optimal $\vec{s}$ to see if a parallel setup is advantageous. We showed in Sec.~\ref{sec:mem} that the optimal number of memories is four using the \DP protocol for $N=3$, $\kvec=(2,0,0,0)$ with a secret key rate per memory per second of $K=0.32$. In Table~\ref{Tab4}, we fixed the number of memories and calculated the optimal key rate by optimizing the remaining parameters. We find that  except for $M=4$, the secret key rate per memory per second is higher (or equal) for the \DurP protocol.
\begin{table}[ht]
\caption{\label{Tab4}Secret key rate per total number of used memories [Eq.~\eqref{eq:KMemtotal}] for the different distillation protocols and for a fixed number of memories $M$. The optimal configurations are given by the distillation vectors $\vec{k}_{M}=(k_0,...,k_N)$, with $\vec{k}_{M}$ denoting the distillation strategy using $M$ memories. The notation $(\kvec_m,\kvec_{m'})$ means parallel setups using $m$ and $m'$ memories. Parameters: $F_0=0.97$ and $p_G=0.99$.}
\begin{ruledtabular}
\begin{tabular}{ccccc}
&\multicolumn{2}{c}{\DurP protocol}&\multicolumn{2}{c}{\DP protocol} \\
$M$& $K$ & configuration &  $K$ &configuration \\\hline
$1$& $0.19$&$\kvec_1=(0,0,0)$& $0.19$& $\kvec_1=(0,0,0)$\\
$2$ &$0.58$ &$\kvec_2=(0,0,2,0)$& $0.58$ & $\kvec_2=(0,0,1,0)$
\\
$3$&$0.96$&$\kvec_3=(0,1,2,0,0)$& $0.45$&$\left(\kvec_1,\kvec_2\right)$ \\
$4$&$0.82$&$\kvec_4=(0,1,1,1,0)$& $0.87$& $\kvec_4=(0,0,2,0,0)$\\
$5$&$0.81$&$\left(\kvec_2,\kvec_3\right)$& $0.73$&$\left(\kvec_1,\kvec_4\right)$\\
$6$&$0.96$& $\left(\kvec_3,\kvec_3\right)$&$0.78$&$\left(\kvec_2,\kvec_4\right)$\\
$7$&$0.89$&$\left(\kvec_3,\kvec_4\right)$&$0.69$&$\left(\kvec_1,\kvec_2,\kvec_4\right)$\\
\end{tabular}
\end{ruledtabular}
\end{table}
\section{\label{sec:Impact}Impact of classical communication on the secret key rate}

In this section we investigate the impact of the classical communication time required for acknowledging the success of entanglement swapping and entanglement distillation on the secret key rate. First we calculate the repeater rates $R_{\text{Rep,NC}}$ where we only consider the classical communication for entanglement distribution. Then we compare the optimal secret key rates using the repeater rate without ($R_{\text{Rep,NC}}$) and with classical communication  ($R_{\text{Rep}}$) [see Eqs.~\eqref{eq:rawDCC} and \eqref{eq:rawDurCC}], and discuss the differences.

The repeater rate for the \DP protocol without the classical communication time due to entanglement swapping and entanglemend distillation is given by (see, e.g., \cite{Sangouard2011,Abruzzo2012})
\begin{equation}
\label{eq:raw}
R_{\rm Rep,NC}^{\rm D}=\frac{1}{2 T_0}\left(\frac{2}{3}\right)^{N+\sum_n k_n}P_0 \prod_{n=1}^NP_{ES}(n)\prod_{i=0}^{k_n} P_D^{\rm D}(i,n),
\end{equation}
which is derived from the solution of the recurrence relation in Eq.~\eqref{allequations} by omitting all terms acknowledging the classical communication time, i.e., the terms with $2^{N-1}$ and $2^{N}$ [see Appendix, Eq.~\eqref{sol:DRepCC}].

The corresponding repeater rate for the \DurP protocol can be derived analogously by omitting terms in the recurrence relation given in Eq.~\eqref{eq:rec}. This leads to 
\begin{equation}
 \label{eq:rawDur}
R_{\rm Rep,NC}^{\text{D\"{u}r}}=\frac{P_0}{2T_0}\left(\frac{2}{3}\right)^N\prod_{i=0}^N\frac{P_{ES}(i)}{a(i)},
\end{equation}
where 
\begin{equation}
a(i)=\prod_{j=0}^{k_i-1}P_D^{\text{ D\"{u}r}}(k_i-j,i)^{-1}\\
+\sum_{m=0}^{k_i-1}\prod_{j=0}^mP_D^{\text{ D\"{u}r}}(k_i-j,i)^{-1}
\end{equation}
and $P_{ES}(0)=1$ (see Appendix~\ref{sec:AppRep} for details).

For investigating the relevance of the classical communication time, we determine the \textit{relative change} of the optimal secret key rates with this classical communication $K(R_{\text{Rep}})$  and without classical communication $K(R_{\rm Rep,NC})$, i.e.,
\begin{equation}
\label{eq:rel1}
\Delta_{\rm rel}(K(R_{\rm Rep,NC}),K(R_{\rm Rep})),
\end{equation}
with $K$ being the optimal secret key rate per memory per second [Eq.~\eqref{eq:KMem}]. The relative change $\Delta_{\rm rel}$ is defined by
\begin{equation}
\label{eq:rel}
\Delta_{\rm rel}(a,b):=\left(a-b\right)/\max\{a,b\}.
\end{equation}
We optimize both secret key rates over the same parameter set as in Sec.~\ref{sec:keyrate}. 

\begin{figure}[!htpb]
\centering
\includegraphics[width=0.32\textwidth, clip, angle=270]{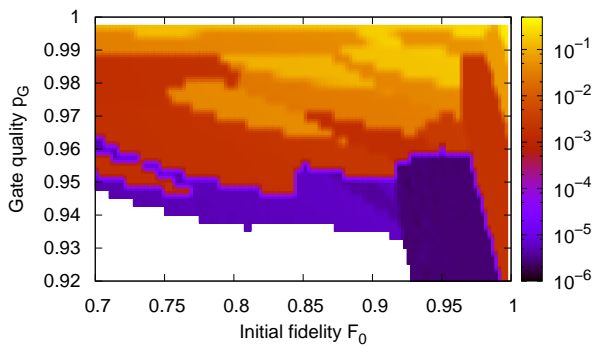}
 \caption{(Color online) The relative change [Eq.~\eqref{eq:rel1}] of the optimal secret key rate per memory [Eq.~\eqref{eq:KMem}] without and with the classical communication time (see text) in terms of the initial fidelity $F_0$ and gate quality $p_G$ ($L=600$ km).}
 \label{fig:OptScrScanWoCC_key_relativeChange}
\end{figure}

Figure~\ref{fig:OptScrScanWoCC_key_relativeChange} shows the relative change of the optimal secret key rate per second per memory.
% , i.e.
% \begin{equation}
% \label{eq:rel2}
%  \Delta_{\rm rel}(K(R_{\rm Rep}),K(R_{\rm Rep,NC})).
% \end{equation}
Depending  on the parameters the secret key rate without the classical communication time $K(R_{\rm Rep,NC})$ can be by a factor of two bigger. This is the yellow region in Fig.~\ref{fig:OptScrScanWoCC_key_relativeChange}. By inspecting Fig.~\ref{fig:multiplot}~(a) the secret key rate in this region is in the order of secret bits per second. Except from some regions, the parameters leading to the optimal secret key rate without and with the classical communication time are almost the same. 

In a previous paper \cite{Dur1999} it was claimed that the main contribution of the entanglement generation time (i.e., the inverse of the repeater rate) is the classical communication time needed for acknowledging the success of entanglement swapping and entanglement distillation. We have seen that this is not the case here. Comparing the results given in Ref.~\cite{Dur1999}, we have found that the relative change [Eq.~\eqref{eq:rel1}] is not more than 40\% and both secret key rates are in the same order of magnitude (distance $L=1280$ km). We discovered that the influence of non-perfect success probabilities for distillation is substantial. Here, the entanglement generation time is one order of magnitude larger than in \cite{Dur1999}, where the success probability of entanglement distillation was not considered (parameters: $F_0=0.96$ and $p_G=0.995$).

Note that here we consider the memories to be perfect. Certainly, if the storage time of the memories  is limited, such an analysis might lead to other results.

\section{\label{sec:conclusion}Conclusion}
For given imperfect initial fidelities and imperfect gates, we found the quantum repeater configurations (i.e., the distillation protocol, distillation strategy, number of distillation rounds, number of nesting levels, and number of memories) that lead to the optimal secret key rate per memory per second. For this purpose we focused on a specific recurrence protocol (\DP) and an entanglement pumping (\DurP) protocol. We found that there exists a regime ($p_G\leq0.99$ and $F_0\leq0.8$) of parameters where the entanglement pumping protocol performs best. However, for lower initial fidelities typically the recurrence protocol is favorable. 

Regarding the distillation strategy [distilling with the same number of rounds in each nesting level (strategy $\alpha$) or distilling only in the beginning (strategy $\beta$)], we have seen that for some parameters, strategy $\beta$, which is experimentally more feasible, is optimal and that this region strongly depends on the detector efficiency.  We found that with decreasing detector efficiency it is optimal to not distill. Lifting the restriction of an equal number of distillation rounds in each nesting level for some set of parameters (initial fidelity and gate quality), we have found that the improvement of the secret rate is not more than one order of magnitude compared to distillation strategy $\alpha$. We also showed that increasing the number of repeater stations and rounds of distillation does not necessarily lead to an increase in the secret key rate.

We investigated the role of the form of the input states, where we used either a depolarized or a binary state. We found that the secret key rate per memory per second for both forms is in the same order of magnitude; the binary states have the advantage that for low fidelities and gate qualities they provide a non-zero secret key rate compared to a depolarized input state. Binary states can be produced by the hybrid quantum repeater.

When fixing the number of memories for a specific set of parameters, we investigated which distillation protocol is optimal, and found that setups working in parallel can be advantageous.

Finally, we derived formulas for the generation rate of entangled pairs per second (\textit{repeater rate}) including the classical communication times for acknowledging the success of entanglement swapping and entanglement distillation. We calculated the secret key rate per memory per second without and with the classical communication time and found that the main contribution is the time to distribute the entangled pairs, which is contrary to the results in the literature.

Further studies could implement the formalism for the quantum repeater in the context of finite keys (see, e.g., \cite{Scarani2009} for a review) and for imperfect memories (see, e.g., \cite{Hartmann2007}). 

\begin{acknowledgments}
The authors acknowledge financial support by the German Federal Ministry of Education and Research (BMBF, project QuOReP).
\end{acknowledgments}

\bibliography{library}
%\onecolumngrid
\appendix*
\section{\label{subsec:rawkey}Solutions for the recurrence formulas}

In this appendix we give the solutions for the recurrence formulas [Eqs.~\eqref{allequations} and \eqref{eq:rec} in Sec.~\ref{sec:rep}] that are needed for calculating the repeater rate for the \DP and the \DurP protocol.

\subsection{The \DP protocol}
We first solve the recurrence relation for Eq.~\eqref{equationa} and terminate when $k_N=0$:
\begin{eqnarray}
\label{sol:DRepb}
 \tauD(k_N,N)&=&\tauD(0,N)\underbrace{\left(\frac{3}{2}\right)^{k_N} \prod_{j=0}^{k_N-1}\frac{1}{P_D^\textrm{D}(k_N-j,N)}}_{=:\alpha(N)}\nonumber\\
&&+\underbrace{2^{N}\sum_{i=0}^{k_N-1}\left(\frac{3}{2}\right)^i\prod_{j=0}^{i}\frac{1}{P_D^\textrm{D}(k_N-j,N)}}_{=:\beta(N)}.
\end{eqnarray}
Then we replace $\tauD(0,N)$ by Eq.~\eqref{equationb} resulting in:
\begin{equation*}
\tauD(k_N,N)=\frac{\alpha(N)}{P_{ES}(N)}\left(\frac{3}{2}\tauD(k_{N-1},N-1)+2^{N-1}\right)+\beta(N),
\end{equation*}
which is another recurrence relation depending on $N$. We can now solve this relation until we reach $\tauD(k_0,0)$:
\begin{eqnarray}
  \tauD(k_N,N)&=&\tauD(k_0,0)\left(\frac{3}{2}\right)^N\prod_{j=0}^{N-1}\frac{\alpha(N-j)}{P_{ES}(N-j)}\nonumber\\
&&+\sum_{i=1}^N\left(\frac{3}{2}\right)^{N-i}2^{i-1} \prod_{j=0}^{N-i}\frac{\alpha(N-j)}{P_{ES}(N-j)}\nonumber\\
&&+\sum_{i=1}^N\left(\frac{3}{2}\right)^{N-i}\beta(i) \prod_{j=0}^{N-(i+1)}\frac{\alpha(N-j)}{P_{ES}(N-j)},\label{sol:DRepCC}
\end{eqnarray}
where we can replace $\tauD(k_0,0)$ by $\tauD(0,0)\alpha(0)+\beta(0)$ using Eq.~\eqref{sol:DRepb}.

\subsection{The \DurP protocol}
\subsubsection{Solution of the recurrence relation Eq.~\eqref{eq:rec}}
The solution of the recurrence relation in Eq.~\eqref{eq:rec} is analogously given by 
\begin{widetext}
\begin{eqnarray}
\label{eq:taukN}
\tauDur(k_N,N)&=&
 \tauDur(0,N)\underbrace{\left(\prod_{j=0}^{k_N-1}P_D^\textrm{D\"{u}r}(k_N-j,N)^{-1}+\sum_{i=0}^{k_N-1}\prod_{j=0}^iP_D^\textrm{D\"{u}r}(k_N-j,N)^{-1}\right)}_{=:a(N)}+\underbrace{2^{N}\left(\sum_{i=0}^{k_N-1}\prod_{j=0}^iP_D^\textrm{D\"{u}r}(k_N-j,N)^{-1}\right)}_{=:b(N)},
\end{eqnarray}
\end{widetext}
where we use the convention that $\sum_{i=0}^{-1}f(i)=0$ and  $\prod_{i=0}^{-1}c(i)=1$.
Inserting now $\tauDur(0,N)=\frac{3}{2}\tauDur(k_{N-1},N-1)+2^{N-1}$ into $\tauDur(k_N,N)=\tauDur(0,N) a(N)+b(N)$ leads to the recurrence relation:
\begin{eqnarray}
\tauDur(k_N,N)&=&\frac{a(N)}{P_{ES}(N)}\left(\frac{3}{2}\tauDur(k_{N-1},N-1)+2^{N-1}\right)\nonumber\\
&&+b(N).
\end{eqnarray}
The solution of this relation is
%\begin{widetext}
\begin{eqnarray}
\label{eq:tauDur2}
 \tauDur(k_N,N)&=&\tau(k_0,0)\left(\frac{3}{2}\right)^N\prod_{j=0}^{N-1}\frac{a(N-j)}{P_{ES}(N-j)}\\ &&+\sum_{i=1}^N\left(\frac{3}{2}\right)^{N-i}2^{i-1} \prod_{j=0}^{N-i}\frac{a(N-j)}{P_{ES}(N-j)}\\
&&+\sum_{i=1}^N\left(\frac{3}{2}\right)^{N-i}b(i) \prod_{j=0}^{N-(i+1)}\frac{a(N-j)}{P_{ES}(N-j)}.
\end{eqnarray}
%\end{widetext}
The solution for $\tauDur(k_0,0)$ we get from Eq.~(\ref{eq:taukN}):
\begin{equation}
\label{eq:tauDur0}
 \tauDur(k_0,0)=\tauDur(0,0)a(0)+b(0).
\end{equation}

\subsubsection{\label{sec:AppRep}Derivation of the repeater rate without the classical communication time for entanglement distillation and entanglement swapping, Eq.~\eqref{eq:rawDur}}
For obtaining the solution for the recurrence relations without classical communication time for entanglement distillation and entanglement swapping, we just set in Eq.~\eqref{eq:taukN} $b(N)=0$. What remains from the solution is just the first term of Eq.~\eqref{eq:tauDur2}, which is exactly 
\begin{equation}
\tauDur_{NC}(k_N,N)=\tauDur_{NC}(k_0,0)\left(\frac{3}{2}\right)^N\prod_{j=0}^{N-1}\frac{a(N-j)}{P_{ES}(N-j)}.
\end{equation}
We replace $\tauDur_{NC}(k_0,0)$ by $\tauDur(0,0)a(0)$ [see Eq.~\eqref{eq:tauDur0}] and get
\begin{equation}
\tauDur_{NC}(k_N,N)=\tauDur(0,0)\left(\frac{3}{2}\right)^N\prod_{j=0}^{N}\frac{a(N-j)}{P_{ES}(N-j)}.
\end{equation}
The repeater rate is given by
\begin{eqnarray}
R_{\rm Rep,NC}^{\text{D\"{u}r}}&=&\frac{1}{T_0 \tauDur_{NC}(k_N,N)}\\
&=&\frac{P_0}{2T_0}\left(\frac{2}{3}\right)^N\prod_{i=0}^N\frac{P_{ES}(i)}{a(i)}, 
\end{eqnarray}
where we used that $\tauDur(0,0)=\frac{2}{P_0}$.

\end{document}